# Antiferromagnetic Tunnel Junctions for Spintronics

Ding-Fu Shao[1,*] and Evgeny Y. Tsymbal [2,†]

[1] *Key Laboratory of Materials Physics, Institute of Solid State Physics, HFIPS, Chinese Academy of Sciences, Hefei 230031, China*

[2] *Department of Physics and Astronomy & Nebraska Center for Materials and Nanoscience, University of Nebraska, Lincoln, Nebraska 68588-0299, USA*

[*] dfshao@issp.ac.cn;  [†] tsymbal@unl.edu

**Abstract** Antiferromagnetic (AFM) spintronics has emerged as a subfield of spintronics, where an AFM Néel vector is used as a state variable. Efficient electric control and detection of the Néel vector are critical for spintronic applications. This review article features fundamental properties of AFM tunnel junctions (AFMTJs) as spintronic devices where such electric control and detection can be realized. We emphasize critical requirements for observing a large tunneling magnetoresistance (TMR) effect in AFMTJs with collinear and noncollinear AFM electrodes, such as a momentum-dependent spin polarization and Néel spin currents. We further discuss spin torques in AFMTJs that are capable of Néel vector switching. Overall, AFMTJs have potential to become a new standard for spintronics providing larger magnetoresistive effects, few orders of magnitude faster switching speed, and much higher packing density than conventional magnetic tunnel junctions (MTJs).

## 1. Introduction

Spintronics is a vigorously developing field of electronics, where electron's spin controls device functionality [1]. Conventional schemes rely on magnetic tunnel junctions (MTJs)—key devices of modern spintronic technologies, such as magnetic random-access memories (MRAMs) (Fig. 1a). In an MRAM, MTJs carry information bits, that can be written-in and read-out by electric means. An important advantage of an MRAM is its non-volatility; however, it is deficient with its low switching speed that is determined by the time required to rotate the magnetization of a ferromagnet, which is typically a few nanoseconds (Fig. 1a). This is nearly three orders of magnitude slower than charging a capacitor in CMOS technologies.

Antiferromagnetic (AFM) spintronics has recently emerged as a subfield of spintronics, where an AFM order parameter known as the Néel vector is used as a state variable [2-4]. Due to being robust against magnetic perturbations, producing no stray fields, and exhibiting ultrafast dynamics, antiferromagnets can serve as promising functional materials for spintronic applications [ 5 ]. Potentially, antiferromagnets can replace ferromagnets due to their orders of magnitude enhanced switching speed and storage density. To realize this potential, efficient electric control and detection of the AFM Néel vector are required. These functionalities can be realized using AFM tunnel junctions (AFMTJs) as core spintronic devices. The recent theoretical predictions [ 6 - 8 ] and experimental demonstrations [9,10] show that AFMTJs can exhibit a strong electric response to the state of the Néel vector, and that the Néel vector itself can be electrically controlled [11]. Potentially, AFM random-access memories (AFM-RAMs) are envisioned (Fig. 1b) that can provide a stronger magnetoresistive response, much faster operation speed, and higher memory density than MRAMs. This review article features fundamental properties of collinear and noncollinear AFMTJs, such as a momentum-dependent spin polarization and Néel spin currents, that control their magnetoresistive properties and discusses spin torques in AFMTJs that are capable of Néel vector switching.

## 2. Magnetic tunnel junctions and tunneling magnetoresistance

A ferromagnet hosts exchange-coupled parallel-aligned magnetic moments carrying a finite magnetization (Fig. 2a), which can be used as a state variable to encode the information. Magnetization is easily controlled by magnetic fields and spin torques, which allows convenient write-in of information. The non-vanishing magnetization in a ferromagnet is inherited from its exchange-split electronic band structure (Figs. 2b and 2c) that is also responsible for a variety of useful spin-dependent transport properties [1]. Among them are those (e.g., magnetoresistance) that allow the electrical detection of the magnetization state for information read-out. To date, most spintronic devices have been based on ferromagnets.

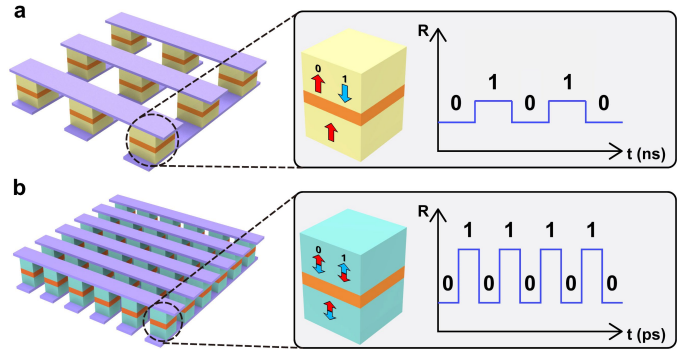

**FIG. 1** (**a**) Schematic of MRAM consisting of an array of conventional MTJs. (**b**) Schematic of AFM-RAM consisting of an array of AFMTJs. AFM-RAMs are expected to exhibit a stronger magnetoresistive response, a much faster operation speed, and higher density compared to the conventional MRAM, due to the advantages of AFMTJs.

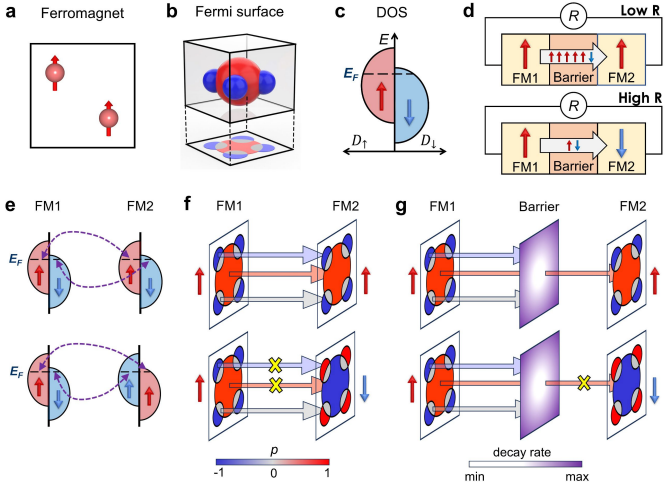

**FIG. 2** (**a-c**) Schematic of magnetic structure (**a**), Fermi surface (**b**), and electronic density of states (DOS) (**c**) of a ferromagnet. Red and blue colors in (**b**) denote up- and down-spin Fermi surfaces. Projection of the Fermi surface into a 2D Brillouin zone represents a distribution of conduction channels. (**d**) Schematic of a conventional MTJ based on two FM electrodes and a nonmagnetic tunnel barrier. (**e**) Schematic of the TMR mechanism based on spin-polarized DOS. (**f**) Schematic of the TMR mechanism based on the momentum-dependent spin-polarization matching of conduction channels in two electrodes. (**g**) Schematic of momentum-dependent spin-filtering in the tunnel barrier.

**Magnetic tunnel junctions (MTJs)** An MTJ is the most common spintronic device utilizing the advantages of ferromagnets. Figure 2d shows schematics of an MTJ, where left and right electrodes are ferromagnetic (FM) metals separated by an insulating non-magnetic tunnel barrier. Magnetization of the left electrode is pinned, while it is free in the right electrode. Parallel (P) and antiparallel (AP) alignments of magnetization in the two electrodes represent "0" and "1" bits of information.

**Tunneling magnetoresistance (TMR)** Electron tunneling in an MTJ can be effectively controlled by the relative orientation of magnetization. Switching between the P and AP states changes the resistance of an MTJ. This effect is known as tunneling magnetoresistance (TMR) and can be used for read-out of "0s" and "1s" in an MTJ [12,13]. The magnitude of the effect is normally quantified by the TMR ratio, $TMR = \frac{R_{AP} - R_P}{R_P}$, where $R_P$ ($R_{AP}$) is resistance of the P (AP) state. TMR can be understood assuming that electron's spin is conserved in the tunneling process, so that tunneling of up- and down-spin electrons occurs in parallel in two spin conduction channels [14]. As a result, TMR can be qualitatively described in terms of Julliere's formula $TMR = \frac{2p_1 p_2}{1 - p_1 p_2}$, where $p_1$ and $p_2$ are spin polarizations of the two electrodes [12]. The widely used definition of the total spin polarization is $p = \frac{D^\uparrow(E_F) - D^\downarrow(E_F)}{D^\uparrow(E_F) + D^\downarrow(E_F)}$, where $D^{\uparrow,\downarrow}(E_F)$ is the spin-dependent density of states (DOS) at the Fermi energy ($E_F$) (Figs. 2c and 2e). Based on this formula, a larger spin polarization of the electrodes favors a larger TMR. While relevant for polycrystalline MTJs, this simple description does not capture anisotropy of the Fermi surface that is essential for tunneling in crystalline MTJs, and it does not reflect effects of the tunnel barrier and interfaces [15].

In crystalline MTJs with no diffuse scattering, tunneling conductance $G$ can be described in the ballistic transport regime where the transverse wave vector $\mathbf{k}_\parallel$ is conserved, so that [16,17]

$$G = \sum_\sigma G^\sigma = \frac{e^2}{(2\pi)^3 \hbar} \sum_\sigma \int T_\parallel^\sigma \, d\mathbf{k}_\parallel, \qquad (1)$$

where $\sigma$ is the spin index, and $T_\parallel^\sigma$ is the $\mathbf{k}_\parallel$- and $\sigma$-dependent transmission. $T_\parallel^\sigma$ is determined by the probability of tunneling of the Bloch states across the barrier at $\mathbf{k}_\parallel$. Due to being separated into $\mathbf{k}_\parallel$-dependent channels conducting in parallel, the spin-dependent conductance in MTJs can be better characterized in terms of the $\mathbf{k}_\parallel$-dependent number of conduction channels $N_\parallel^\sigma(\mathbf{k}_\parallel)$ in the electrodes, rather than the total DOS—a characteristic integrated over $\mathbf{k}_\parallel$. The number of conduction channels is defined by the number of propagating Bloch states in the transport direction at the Fermi energy [18]:

$$N_\parallel^\sigma(\mathbf{k}_\parallel) = \frac{\hbar}{2} \sum_n \int |v_{nz}^\sigma| \frac{\partial f}{\partial E_n^\sigma(\vec{k})} dk_z, \qquad (2)$$

where $E_n^\sigma(\mathbf{k})$ is energy of the $n$-th band, $v_{nz}^\sigma = \frac{\partial E_n^\sigma(\mathbf{k})}{\hbar \partial k_z}$ is the band velocity along the transport $z$ direction, and $f$ is the Fermi distribution function. A $\mathbf{k}_\parallel$-dependent spin polarization $p_\parallel$ can then be defined by

$$p_\parallel(\mathbf{k}_\parallel) = \frac{N_\parallel^\uparrow - N_\parallel^\downarrow}{N_\parallel^\uparrow + N_\parallel^\downarrow}, \qquad (3)$$

and the net spin polarization by

$$p = \frac{\sum_{\mathbf{k}_\parallel} (N_\parallel^\uparrow - N_\parallel^\downarrow)}{\sum_{\mathbf{k}_\parallel} (N_\parallel^\uparrow + N_\parallel^\downarrow)}. \qquad (4)$$

These quantities capture electron's spin and velocity at the Fermi surface, and thus are more relevant for the description of transport properties than the DOS. They can be regarded as *transport spin polarizations* of a magnetic metal in the ballistic transport regime. In crystalline MTJs, the $\mathbf{k}_\parallel$-dependent transport spin polarization $p_\parallel$ is more appropriate to quantify TMR than the net spin polarization $p$. Based on $p_\parallel$, TMR in crystalline MTJs can be explained by stronger transmission for a P magnetization state than for an AP state due to the matching of $p_{\parallel,1}$ and $p_{\parallel,2}$ in the two electrodes (labeled by 1 and 2) for the P state and mismatching for the AP state (Fig. 2e).

The tunneling barrier also plays an important role in TMR. First, transmission is expected to be stronger at those $k_\parallel$ where the decay rate in the barrier is lower. Second, barrier effectively transmits only those Bloch states that have symmetry matching to the low-decay-rate evanescent states in the barrier [19,20]. This process may significantly enhance TMR if the barrier selects conduction channels with a high degree of spin polarization (Fig. 2f). For example, the matching of the majority-spin $\Delta_1$ band in the Fe (001) electrode to the $\Delta_1$ evanescent state in the MgO (001) barrier is responsible for a large positive spin polarization and giant values of TMR predicted [20, 21] and observed [22, 23] in crystalline Fe/MgO/Fe (001) MTJs.

## 3. AFMTJs with collinear AFM electrodes

A vanishing net magnetization of antiferromagnets (Figs. 3a and 3b) makes the realization of TMR in AFMTJs much more challenging. AFMTJs represent tunnel junctions with two AFM electrodes whose Néel vector can be aligned parallel (P) or antiparallel (AP) resulting in the TMR effect (Fig. 3c). Various approaches have been theoretically proposed to observe TMR in AFMTJs, such as utilizing the quantum coherence between the staggered spin densities in AFM electrodes or between two uncompensated interfaces [24-29]. These effects, however, are not robust against interface roughness and disorder [30,31], and likely because of these reasons have not been confirmed in experiment. Recently, it has been predicted that some collinear antiferromagnets can exhibit momentum-dependent [32-41] and sublattice-dependent [11] transport spin polarization, even in the absence of spin-orbit coupling (SOC). These collinear antiferromagnets appear to be promising materials to serve as electrodes in AFMTJs [6,8,11].

**TMR due to momentum-dependent spin polarization** An antiferromagnet is a magnetically ordered material with equivalent magnetic moments exactly compensated, and thus zero net spontaneous magnetization [42, 43]. In collinear antiferromagnets with two antiparallel aligned magnetic sublattices, there exists a symmetry $\hat{O}$ enforcing $\boldsymbol{m}_A = -\boldsymbol{m}_B$, where $\boldsymbol{m}_\alpha$ is the magnetization on magnetic sublattice $\alpha = A, B$ (Figs. 3a and 3b). The Néel vector $\boldsymbol{n} = \boldsymbol{m}_A - \boldsymbol{m}_B$ can serve as the magnetic order parameter. The most common $\hat{O}$ symmetry is $\hat{P}\hat{T}$ that combines space inversion $\hat{P}$ and time reversal $\hat{T}$ (Fig. 3a). This symmetry enforces Kramers' spin degeneracy in the band structure, as $\hat{P}\hat{T}E_n^\uparrow(\boldsymbol{k}) = E_n^\downarrow(\boldsymbol{k})$ (Fig. 3a). This spin degeneracy also appears in compensated antiferromagnets with $\hat{O} = \hat{U}\hat{t}$ symmetry ($\hat{U}$ is spin rotation and $\hat{t}$ is half a unit cell translation) in the absence of SOC. As a result, the spin polarization $p_\parallel$ of all conduction channels vanishes (Fig. 3a).

The magnetization compensation in an antiferromagnet can also be ensured by symmetries other than $\hat{P}\hat{T}$ and $\hat{U}\hat{t}$. Mirror/glide planes and rotation/screw axes in a crystal can also enforce $\boldsymbol{m}_A = -\boldsymbol{m}_B$. In such compensated antiferromagnet with $\hat{P}\hat{T}$ and $\hat{U}\hat{t}$ symmetries broken, spin splitting of the band structure appears even in the absence of SOC. This subgroup of antiferromagnets was dubbed as "altermagnets" [39,40] and their spin-split electronic structure was well understood based on the symmetry analyses [32-41]. Particularly, the spin point/space group theories [44,45] provide powerful tools to understand the nonrelativistic properties of magnets and their relation to spintronics [38,39, 46-49].

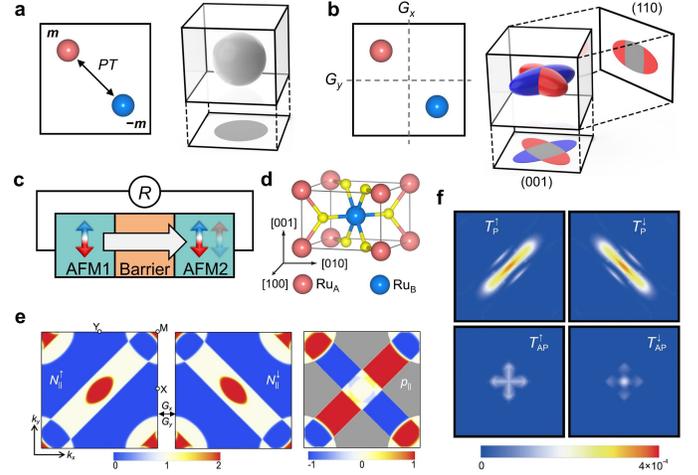

**FIG. 3** (**a**) Schematics of the magnetic structure and the Fermi surface of a collinear antiferromagnet with magnetization compensated by $\hat{P}\hat{T}$ symmetry. The symmetry enforces spin degeneracy of the Fermi surface and conduction channels (indicated by grey color). (**b**) Schematics of the magnetic structure and the Fermi surface of a collinear antiferromagnet with magnetization compensated by two glide symmetries $\hat{G}_x$ and $\hat{G}_y$. The symmetries allow spin splitting at wavevectors away from the $\hat{G}_x$ and $\hat{G}_y$ invariant planes (indicated by red and blue colors). This leads to the momentum-dependent spin polarization that is compensated in the (001) plane and uncompensated in the (110) plane. (**c**) Schematic of an AFMTJ with two AFM electrodes and a nonmagnetic tunnel barrier. (**d**) The atomic structure of a collinear AFM $RuO_2$, which hosts two sublattices $Ru_A$ and $Ru_B$ with antiparallel magnetic moments. (**e**) Momentum-dependent conduction channels and associated spin polarizations in the 2D Brillouin zone of $RuO_2$ (001). (**f**) Calculated momentum-dependent transmission of a $RuO_2/TiO_2/RuO_2$ (001) AFMTJ. Figures (**c-f**) are reprinted from Ref. [6] under permission of the Creative Commons CC BY license.

Figure 3b shows a typical spin-split Fermi surface of an altermagnet. Here, although $\hat{P}\hat{T}$ and $\hat{U}\hat{t}$ symmetries are broken, there are two glide symmetries, $\hat{G}_x$ and $\hat{G}_y$, that connect two sublattices. The Fermi surface exhibits an anisotropic spin distribution $s(\boldsymbol{k})$ that is $\hat{T}$-odd, i.e. $s(\boldsymbol{k}) = s(-\boldsymbol{k})$. This contrasts with a $\hat{T}$-even spin distribution in SOC-split non-magnets, where $s(\boldsymbol{k}) = -s(-\boldsymbol{k})$. The anisotropic $\hat{T}$-odd $s(\boldsymbol{k})$ makes transport properties direction dependent. For example, in the [001] transport direction, the $\boldsymbol{k}_\parallel$-dependent conduction channels in the (001) 2D Brillouin zone host antisymmetric spin polarizations, i.e. $p_\parallel(k_x, k_y) = -p_\parallel(k_x, -k_y) = -p_\parallel(-k_x, k_y)$, due to the [001] direction being invariant under $\hat{G}_x$ and $\hat{G}_y$. As a result, a current flowing along the [001] direction is globally spin neutral [6]. On the contrary, an uncompensated $p_\parallel$ appears in the conduction channels in the (110) 2D Brillouin zone, due to the [110] direction not being invariant under $\hat{G}_x$ and $\hat{G}_y$. This allows a net spin-polarized current along the [110] direction with a finite transport spin polarization $p$ (Fig. 3b) [50, 51].

Due to TMR being controlled by the $\bm{k}_\parallel$-dependent spin polarization $p_\parallel$, even for the transport direction supporting only spin-neutral currents ($p = 0$), altermagnets can produce TMR in AFMTJs. This possibility has been investigated for $RuO_2$ [6], a high-temperature AFM metal discovered recently [52, 53]. $RuO_2$ has a rutile structure with two magnetic sublattices $Ru_A$ and $Ru_B$ (Fig. 3d). Its magnetic space group $P4_2'/mnm'$ ensures the compensated magnetization and spin-split electronic structure. The $RuO_2$ Fermi surface has similar characteristics to those shown in Fig. 3b. In the (001) stacking, the conduction channels reveal an antisymmetric distribution of $p_\parallel$ with respect to the $\hat{G}_x$ and $\hat{G}_y$ planes in the 2D Brillouin zone (Fig. 3e). Since switching the AFM Néel vector reverses $p_\parallel$, matching $p_\parallel$ in a $RuO_2$ (001)-based AFMTJ can be controlled by the relative orientation of the Néel vector in the two electrodes. This ensures a finite TMR. First-principles quantum-transport calculations for all-rutile $RuO_2/TiO_2/RuO_2$ (001) AFMTJs confirm this prediction [6]. As seen from Fig. 3f, the distribution of $T_\parallel^\sigma$ in the P state echoes the distribution of $N_\parallel^\sigma$ in bulk $RuO_2$ (001), while $T_\parallel^\sigma$ in the AP state is blocked at the wavevectors with $|p_\parallel| = 1$ in bulk $RuO_2$ (001). The resulting TMR is ~500% which is comparable to the TMR predicted for conventional Fe/MgO/Fe MTJs [20,21].

A giant TMR also appears in a $RuO_2/TiO_2/RuO_2$ (110) AFMTJ, where $p_\parallel$ is uncompensated in the $RuO_2$ (110) 2D Brillouin zone supporting a spin-polarized current with a finite $p$ [51]. While in this case, TMR is expected directly from the presence of the net spin polarization $p$ of $RuO_2$ (110) (like in a FM MTJ), this conventional contribution to TMR appears to be small compared to the contribution associated with the matching of $\bm{k}_\parallel$-dependent spin polarization $p_\parallel$ in the two electrodes.

Altermagnets such as $RuO_2$ (110) can also serve as a counter electrode in MTJs with a single FM electrode. Since both FM and AFM electrodes have finite and uncompensated $p_\parallel$, the TMR is expected to occur due to the $p_\parallel$ matching mechanism. This approach can be used to verify the application potential of altermagnets, since FM electrodes can be easily switched by an applied magnetic field. From the practical perspective, this can also simplify the design of a conventional MTJ, due to no need for an additional pinning layer. The giant TMR of $RuO_2/TiO_2/CrO_2$ (110) all-rutile MTJ has been predicted recently [54,55], using half-metallic $CrO_2$ [56,57] as a FM electrode and $RuO_2$ as an AFM counter electrode.

**TMR due to Néel spin currents** In addition to the momentum-dependent spin polarization, the sublattice-dependent spin polarization in real space can also result in TMR in collinear AFMTJs [11]. In a collinear AFM metal with two magnetic sublattices $\bm{m}_A$ and $\bm{m}_B$, a longitudinal charge current $J$ can be decomposed into intra-sublattice current $J_{\alpha\alpha}$ ($\alpha = A, B$) and inter-sublattice current $J_{\alpha\beta}$ ($\alpha \neq \beta$) (Fig. 4a), such that

$$J_{\alpha\beta} = J_{\alpha\beta}^\uparrow + J_{\alpha\beta}^\downarrow. \tag{5}$$

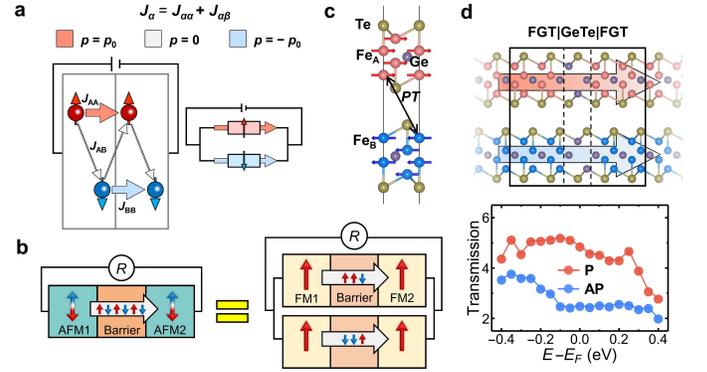

**FIG. 4** (**a**) Schematics of staggered Néel spin currents in a collinear antiferromagnet with strong intra-sublattice coupling. (**b**) AFMTJ with Néel spin currents that can be qualitatively considered as two MTJs connected in parallel. (**c**) $\hat{P}\hat{T}$ symmetric 2D A-type AFM metal $Fe_4GeTe_2$ that supports Néel spin currents. (**d**) Lateral AFMTJ with sizable TMR despite the spin-degenerate electronic structure of $Fe_4GeTe_2$ electrodes. The figures are reprinted from Ref. [11] with permission.

The associated spin current $J_{\alpha\beta}^s$ is

$$J_{\alpha\beta}^s = J_{\alpha\beta}^\uparrow - J_{\alpha\beta}^\downarrow. \tag{6}$$

The spin current flowing through sublattice $\alpha$, dubbed the Néel spin current, I given by

$$J_\alpha^s = \sum_\beta J_{\alpha\beta}^s. \tag{7}$$

It hosts a sublattice-dependent spin polarization

$$p_\alpha = \frac{\sum_\beta J_{\alpha\beta}^s}{\sum_\beta J_{\alpha\beta}}. \tag{8}$$

The intra- and inter-sublattice currents are determined by the inter- and intra-sublattice electron hopping along the transport direction. Since the intra-sublattice currents $J_{AA}$ and $J_{BB}$ are spin-polarized, the Néel spin currents with large $p_\alpha$ can emerge if the intra-sublattice hopping is dominant.

For A-type antiferromagnets composed of antiparallel-aligned FM layers and C-type antiferromagnets composed of antiparallel-aligned chains, the intra-sublattice electron hopping is usually stronger than the inter-sublattice hopping. In these cases, the two AFM sublattices can be considered as connected in parallel with staggered Néel spin currents on the sublattices (Fig. 4a). AFMTJs based on such AFM electrodes can then be qualitatively considered as two MTJs connected in parallel, which naturally supports TMR (Fig. 4b).

The Néel spin currents do not rely on spin-split electronic structure, and hence can emerge even in $\hat{P}\hat{T}$ symmetric antiferromagnets with Kramers' spin degeneracy. For example, in the recently discovered two-dimensional (2D) van der Waals magnet Fe$_4$GeTe$_2$ (Fig. 4c), where the AFM order is induced by doping [58], Néel spin currents with a large spin polarization $|p_\alpha| = 68\%$ for each layer are predicted, despite the spin-degenerate band structure (Fig. 4d), resulting in a sizable TMR in a Fe$_4$GeTe$_2$-based AFMTJ (Fig. 4e) [11]. Such a lateral junction can be realized experimentally using the recently developed edge-epitaxy technique [59-61].

The TMR in the RuO$_2$/TiO$_2$/RuO$_2$ (001) AFMTJ discussed above can be also understood in terms of the Néel spin currents [11]. This is due to rutile $M$O$_2$ ($M$ is a transition metal element) being composed of edge-sharing $M$O$_6$ octahedra chains along the [001] direction, where the adjacent chains share common corners of the octahedra. Therefore, RuO$_2$ can be regarded as a C-type antiferromagnet that supports Néel spin currents with a non-zero spin polarization. This fact has been verified for a RuO$_2$/TiO$_2$/[TiO$_2$/CrO$_2$]$_n$/CrO$_2$ (001) MTJ, where [TiO$_2$/CrO$_2$]$_n$ represents a multilayer of alternating TiO$_2$ (001) and CrO$_2$ (001) monolayers with $n$ repeats [54]. The latter can be fabricated using modern thin-film growth techniques [62,63]. Although $p_\parallel$ matching of bulk RuO$_2$ (001) and CrO$_2$ (001) electrodes is unable to generate TMR by symmetry, the presence of the [TiO$_2$/CrO$_2$]$_n$ multilayer results in a different effective barrier thickness for the Néel spin currents flowing on Ru$_A$ and Ru$_B$ sublattices. As a result, this AFMTJ can be decomposed into two parallel-connected MTJs with different barrier thickness, where conduction is dominated by the MTJ with smaller barrier thickness. This generates sizable TMR and proves the existence of the Néel spin currents [54].

## 4. AFMTJs with noncollinear AFM electrodes

In a magnetically frustrated crystal structure, collinear magnetic moment alignment may not guarantee the lowest energy. For example, in a Kagome lattice with AFM nearest-neighbor exchange interactions, the co-planar moments form a noncollinear AFM alignment with a 120° angle between each other (Fig. 5a). The Néel vector for such antiferromagnets is not uniquely defined, and the noncollinear AFM order is often represented by a magnetic multipole, a magnetic toroidal multipole [64], or even by direction of a small net magnetization generated by magnetic moment canting due to SOC [65,66]. The latter allows these noncollinear antiferromagnets to be considered as weak ferromagnets exhibiting spin-dependent transport properties, such as the anomalous Hall effect [67-69]. This fact has stimulated broad interest in the properties of noncollinear antiferromagnets.

**Spin polarization in noncollinear antiferromagnets** Noncollinear antiferromagnets exhibit lower symmetry compared to their collinear counterparts. As a result, they generally support nonrelativistic spin-split band structures and spin-textured Fermi surfaces (Fig. 5b) [70, 71, 72], even in the presence of $\hat{U}\hat{t}$ symmetry [47]. Different from $\hat{T}$-even spin textures in nonmagnetic materials induced by SOC, the $\hat{T}$-odd spin textures support longitudinal spin-polarized currents [70, 73], indicating the possibility of using them as electrodes in AFMTJs (Fig. 5c).

However, since spin is not a good quantum number in noncollinear antiferromagnets, it is not clear if the spin-matching mechanism is still valid for noncollinear AFMTJs. The definition of $p_\parallel$ given by Eq. (3) is inappropriate in this case. One can redefine the $\mathbf{k}_\parallel$-dependent spin polarization as a vector [74]

$$\mathbf{p}_\parallel(\mathbf{k}_\parallel) = \frac{\mathbf{s}_\parallel}{\sum_n |\mathbf{s}_{\parallel,n}|}, \qquad (9)$$

where $\mathbf{s}_{\parallel,n}$ is the spin expectation value for band $n$ at $\mathbf{k}_\parallel$ and $E_F$, and $\mathbf{s}_{\parallel,i}$ is the net spin $\mathbf{s}_\parallel = \sum_n \mathbf{s}_{\parallel,n}$ at $\mathbf{k}_\parallel$. This definition is equivalent to Eq. (3) for the collinear case. For noncollinear magnets, although the magnitudes and orientations of $\mathbf{s}_\parallel$ vary with $\mathbf{k}_\parallel$, the magnitude of spin polarization $p_\parallel = |\mathbf{p}_\parallel|$ is large when spins $\mathbf{s}_{\parallel,n}$ at $\mathbf{k}_\parallel$ are nearly parallel in all conduction channels $n$, and is exactly 100% when only one conduction channel is present. Therefore, for noncollinear AFMTJs with two identical electrodes, the spin matching mechanism is expected to work if $p_\parallel$ is large [74].

**TMR in noncollinear AFMTJs** The possibility of TMR in noncollinear AFMTJs has been proposed based on the predicted transport spin polarizations of Mn$_3X$ ($X$ = Sn and Ir) [70] and $A$NMn$_3$ ($A$ = Ga, Ni, Sn, or Pt) [73], and then calculated from first principles for AFMTJs with noncollinear AFM Mn$_3$Sn electrodes [7]. Mn$_3$Sn has a hexagonal D0$_{19}$ structure of space group $P6_3$/mmc, where Mn atoms form a Kagome-type frustrated lattice, with the magnetic moments aligned with 120° angles between each other [66] (Fig. 5d). Such a magnetic alignment belongs to a magnetic space group $Cmc'm'$ that has $\hat{P}\hat{T}$ and $\hat{U}\hat{t}$ symmetries broken. This results in a non-spin-degenerate electronic structure with three Fermi surface sheets, each having finite momentum-dependent spin expectation values. Figure 5e shows the spin texture contributed by one of these sheets, indicating a finite $\mathbf{p}_\parallel$ and thus a possibility for Mn$_3$Sn to serve as electrodes in a noncollinear AFMTJ. There are three other equivalent magnetic states in Mn$_3$Sn, which can be obtained by a 60° rotation around the [001] axis. Switching between these magnetic alignments in one Mn$_3$Sn electrode while keeping the other fixed in a Mn$_3$Sn/vacuum/Mn$_3$Sn AFMTJ changes the matching conditions for $\mathbf{p}_\parallel$ and generates sizable TMR as large as ~300% (Fig. 5f) [7]. A large TMR has also been predicted for noncollinear AFMTJs based on $Mn_3Pt$ [9] and $GaNMn_3$ [74] electrodes.

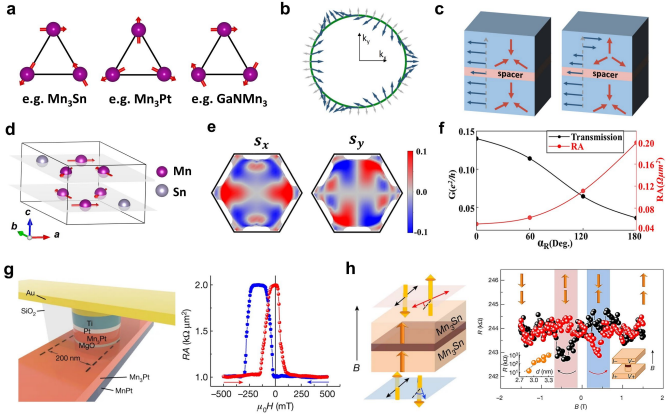

**FIG. 5** (**a**) Schematic of the typical noncollinear AFM alignments. (**b**) Non-relativistic anisotropic spin texture at the Fermi surface of a noncollinear antiferromagnet supporting spin-polarized currents. (**c**) Schematic of a noncollinear AFMTJ with P (left) and AP (right) Néel vectors. Figures (**b**) and (**c**) are reprinted from Ref. [70] with permission. (**d**) Atomic and magnetic structure of $Mn_3Sn$. (**e**) Projected to the (0001) plane spin distribution on a selected Fermi surface sheet in $Mn_3Sn$. (**f**) Calculated tunneling conductance $G$ per lateral unit cell area (left axis) and resistance-area ($RA$) product (right axis) for a $Mn_3Sn$/vacuum/$Mn_3Sn$ AFMTJ when the magnetic domain shown in (**d**) is rotated by an angle $\alpha_R$ around the [0001] axis. Figures (**e**,**f**) are reprinted from Ref. [7] with permission. (**g**,**h**) Experimental results on TMR in $Mn_3Pt$/MgO/$Mn_3Pt$ (**g**) and $Mn_3Sn$/MgO/$Mn_3Sn$ (**h**) tunnel junctions. Left panels schematically show geometry of the AFMTJs; right panels show resistance vs applied magnetic field. Figure (**g**) is reprinted from Ref. [9] with permission. Figure (**h**) is reprinted from Ref. [10] under permission of the Creative Commons CC BY license.

Unlike collinear antiferromagnets whose magnetic state is difficult to control by external stimuli (this is why collinear AFMTJs have not been experimentally realized yet), noncollinear antiferromagnets can be controlled by a magnetic field [66,75] or a spin torque [76-79]. This is due to nonvanishing net magnetization induced by SOC [66], strain [80], or interfaces [81]. Therefore, a noncollinear AFTMJ is easier to realize in experiment.

Recently, two independent experiments have been successfully performed to observe TMR in noncollinear AFMTJs [9, 10]. The first one [9] utilized cubic AFM $Mn_3Pt$ as electrodes and MgO as barrier in this AFMTJ (Fig. 5g). The bottom $Mn_3Pt$ (001) layer was pinned by the exchange bias from an adjacent collinear AFM MnPt layer, while the top $Mn_3Pt$ layer was free to be switched by an external magnetic field. The maximum TMR in this AFMTJ was found to be about 100% at room temperature (Fig. 5g) and about 138% at 10 K. Several devices have been tested and more than 50% of them had room-temperature TMR of more than 70%.

Another work [10] studied noncollinear $Mn_3Sn$/MgO/$Mn_3Sn$ AFMTJ (Fig. 5h), where the bottom epitaxial $Mn_3Sn$ layer had a $(01\bar{1}1)$ orientation, and the top $Mn_3Sn$ layer was polycrystalline. The P and AP states in this AFMTJ could be switched by the magnetic field producing TMR of about 2% at room temperature (Fig. 5h). The effect appeared to be not as large as that predicted [7] likely due to polycrystallinity of the top magnetic layer.

## 5. Spin torques in AFMTJs

Although magnetic fields are widely used to toggle between P and AP states in MTJs, their generation requires substantial currents making them energy inefficient. For low-power and high-density applications, spin torques are more favorable and thus have been extensively explored [82-88]. The spin-torque control of magnetization is even more critical for AFMTJs, since most AFM electrodes (except those with SOC-induced small net magnetic moment), are insensitive to a magnetic field.

**Spin torques for magnetic switching** The dynamics of a magnet can be well described by the Landau-Lifshitz-Gilbert-Slonczewski equation [82,89]:

$$\dot{\boldsymbol{m}}_\alpha = -\gamma \boldsymbol{m}_\alpha \times \boldsymbol{H}_{eff,\alpha} + \alpha_G \boldsymbol{m}_\alpha \times \dot{\boldsymbol{m}}_\alpha + \boldsymbol{\tau}_\alpha^{FL} + \boldsymbol{\tau}_\alpha^{DL}, \quad (10)$$

where $\boldsymbol{m}_\alpha$ is the magnetization of sublattice $\alpha$, $\alpha_G$ is the damping constant, $\gamma$ is the gyromagnetic ratio. The first two terms in Eq. (10) describe the precession and damping torques induced by the intrinsic effective field $\boldsymbol{H}_{eff,\alpha}$. The two last terms are external field-like and damping-like spin torques $\boldsymbol{\tau}_\alpha^{FL} \propto -\boldsymbol{m}_\alpha \times \boldsymbol{p}_\alpha$ and $\boldsymbol{\tau}_\alpha^{DL} \propto -\boldsymbol{m}_\alpha \times (\boldsymbol{m}_\alpha \times \boldsymbol{p}_\alpha)$, where $\boldsymbol{p}_\alpha$ is the current induced non-equilibrium spin polarization on sublattice $\boldsymbol{m}_\alpha$. For ferromagnets, $\boldsymbol{H}_{eff,\alpha}$ is mostly determined by the anisotropy field $\boldsymbol{H}_{K,\alpha}$, and spin torques driven by $\boldsymbol{p}_\alpha$ can directly compete with the intrinsic torques to switch magnetization. The switching of ferromagnets is therefore controlled by $\boldsymbol{H}_{K,\alpha}$ and occurs in the GHz frequency range.

In MTJs, spin polarization $\boldsymbol{p}_\alpha$ is carried by the longitudinal current flowing across the junction. The tunneling current emitted by one electrode carries spin angular momentum that can be transferred to the magnetic moment of the other electrode, generating a spin-transfer torque (STT) [84]. In addition, spin torques can be produced by an in-plane current via spin-Hall [87] and Rashba-Edelstein effects [85]. This requires an additional spin-source layer that is adjacent to the free layer and has a large SOC. This type of spin torque is known as the spin-orbit torque (SOT).

**Spin torques in collinear AFM electrodes** Switching antiferromagnets by spin torques is more complicated than switching ferromagnets [88,90-96]. For example, in a simple collinear antiferromagnet with two antiparallel sublattices ($\alpha = A, B$), $\boldsymbol{H}_{eff,\alpha}$ contains an additional exchange field $\boldsymbol{H}_{E,\alpha}$, which exerts exchange torques $\boldsymbol{\tau}_{E,\alpha}^F \propto -\boldsymbol{m}_\alpha \times \boldsymbol{H}_E$ and $\boldsymbol{\tau}_{E,\alpha}^D \propto -\boldsymbol{m}_\alpha \times$ $(\boldsymbol{m}_\alpha \times \boldsymbol{H}_{eff,\alpha})$. Here, superscripts $F$ and $D$ denote the associated precession (field) and damping torques, respectively. Since $\boldsymbol{H}_{E,\alpha} \propto -\boldsymbol{m}_\beta$ ($\alpha \neq \beta$), $\boldsymbol{\tau}_{E,\alpha}^F$ and $\boldsymbol{\tau}_{E,\alpha}^D$ vanish if the two sublattices are antiparallel. However, if $\boldsymbol{m}_\alpha$ and $\boldsymbol{m}_\beta$ are tilted due to a spin torque, the exchange field generates staggered $\boldsymbol{\tau}_{E,\alpha}^F$. Due to $\boldsymbol{H}_{E,\alpha}$ being typically a factor of $10^3$ greater than $\boldsymbol{H}_{K,\alpha}$, it is practically impossible to create such a large spin torque to directly compete with $\boldsymbol{\tau}_{E,\alpha}^F$ for switching the Néel vector. However, the spin torque can be used to tilt $\boldsymbol{m}_\alpha$ and $\boldsymbol{m}_\beta$ enforcing a Néel vector precession driven by $\boldsymbol{\tau}_{E,\alpha}^F$ [90]. Due to the large $\boldsymbol{H}_{E,\alpha}$, such precession occurs in the THz frequency range. This property can be used for the switching of the Néel vector, provided that an appropriate torque can be generated.

We, first, consider a damping-like torque that is usually employed to switch ferromagnets and can be generated by a uniform spin polarization $\boldsymbol{p}$ of a tunneling current, a spin-Hall effect, or a Rashba-Edelstein effect [82-88]. Normally, in collinear antiferromagnets, these effects lead to $\boldsymbol{p}_A = \boldsymbol{p}_B = \boldsymbol{p}$ for the two sublattices, which results in a uniform damping-like torque $\boldsymbol{\tau}_A^{DL} = \boldsymbol{\tau}_B^{DL}$ and staggered field-like torque $\boldsymbol{\tau}_A^{FL} = -\boldsymbol{\tau}_B^{FL}$.

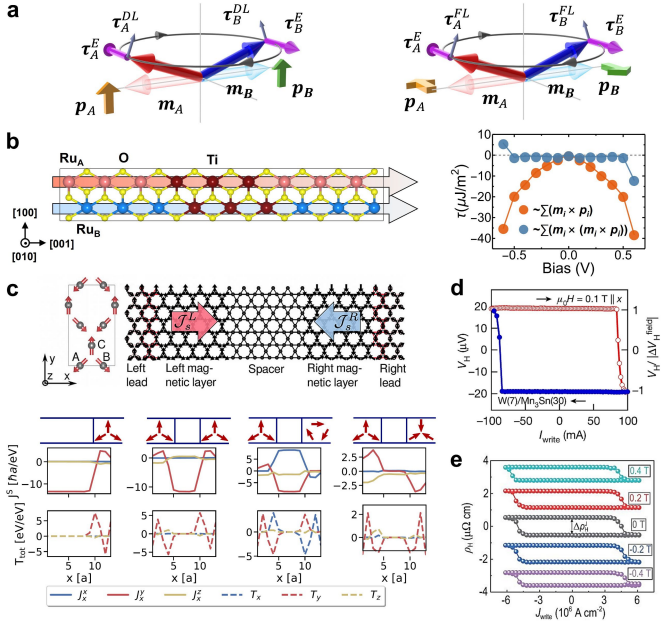

**FIG. 6** (**a**) Schematics of the dynamics of collinear antiferromagnets induced by a spin current with uniform spin polarizations (left) and staggered spin polarizations (right) on two sublattices. (**b**) Atomic structure of a RuO$_2$/TiO$_2$/RuO$_2$ (001) AFMTJ (left) and calculated STT driven by Néel spin currents in this AFMTJ (right). Reprinted from Ref. [11] with permission. (**c**) Schematic noncollinear AFMTJ (top) and calculated STT in this AFMTJ for different magnetic states of the electrodes (bottom). Reprinted from Ref. [96] with permission. (**d**) SOT switching of an epitaxial Mn$_3$Sn film with assistance of a magnetic field. Reprinted from Ref. [77] with permission. (**e**) Field-free spin-torque switching of a Mn$_3$Sn polycrystalline film. Reprinted from Ref. [79] under permission of the Creative Commons CC BY license.

The former tilts $\boldsymbol{m}_A$ and $\boldsymbol{m}_B$ causing the staggered $\boldsymbol{\tau}_{E,\alpha}^F$ to drive an ultrafast oscillation of the Néel vector in the plane perpendicular to $\boldsymbol{p}$ (Fig. 6a, left). This oscillation is persistent that is promising for THz applications [90]. However, the damping-like spin torque induced by a uniform polarization $\boldsymbol{p}$ is *indeterministic* for the reversal of the Néel vector.

We note, however, that such spin torque can be used to rotate the Néel vector between the easy axes in antiferromagnets with multi-axial anisotropy. For example, in an antiferromagnet with bi-axial anisotropy and easy axes along the $x$ and $y$ directions, the Néel vector can be switched from the initial $x$ direction to the final $y$ (or $-y$) direction. This is because a current-induced uniform $\boldsymbol{p}$ along the $x$ axis exerts a damping-like spin torque that drives a persistent oscillation confined within the $y$-$z$ plane. When the current is released, the Néel vector relaxes to the easy $y$ axis. This approach has been employed in heterostructures of antiferromagnet/heavy metal bilayers, where $\boldsymbol{p}$ was generated by the spin Hall effect in the heavy metal layers [95,97-99].

In contrast, if a staggered spin polarization is induced in an antiferromagnet, i.e. $\boldsymbol{p}_A = -\boldsymbol{p}_B$, the staggered damping-like torque $\boldsymbol{\tau}_A^{DL} = -\boldsymbol{\tau}_B^{DL}$ and the uniform field-like torque $\boldsymbol{\tau}_A^{FL} = \boldsymbol{\tau}_B^{FL}$ can be generated [93,94]. The latter tilts $\boldsymbol{m}_A$ and $\boldsymbol{m}_B$, resulting in the staggered $\boldsymbol{\tau}_{E,\alpha}^F$ that drives the rotation of $\boldsymbol{m}_\alpha$ toward the direction of $\boldsymbol{p}_\alpha$ (Fig. 6a, right). When $\boldsymbol{m}_\alpha$ is rotated to be parallel to $\boldsymbol{p}_\alpha$, $\boldsymbol{\tau}_\alpha^{FL}$ vanishes and the dynamics of the Néel vector stops. Therefore, if the staggered $\boldsymbol{p}_\alpha$ is collinear to the easy axis of an antiferromagnet, the ultrafast and *deterministic* switching of the Néel vector can be achieved.

Previously, such a staggered $\boldsymbol{p}_\alpha$ and uniform $\boldsymbol{\tau}_\alpha^{FL}$ have been considered in AFMTJs based on quantum coherence between the staggered spin densities in the AFM electrodes or between two interfaces [24-29]. These effects, however, are not robust against disorder and suffer strong spin dephasing [29-31]. It was found that $\hat{P}\hat{T}$-symmetric antiferromagnets can support staggered $\boldsymbol{p}_\alpha$ via the Rashba-Edelstein effect [93,94], provided that the sublattices in such antiferromagnets are locally non-centrosymmetric [100,101]. This generates the Néel SOTs that have been confirmed experimentally [94, 102]. However, antiferromagnets with this property, such as CuMnAs [94], Mn$_2$Au [93,102], and MnPd$_2$ [103], support neither the momentum-dependent spin polarizations nor notable Néel spin currents, and thus are not suitable to serve as electrodes in AFMTJs.

The required staggered $\boldsymbol{p}_\alpha$ and uniform $\boldsymbol{\tau}_\alpha^{FL}$ can be realized in AFMTJs due to the Néel spin currents. Tunneling Néel spin currents can transfer the staggered $\boldsymbol{p}_\alpha$ from the reference layer to the free layer, exerting the required torques for switching. First-principles quantum-transport calculations have been performed for RuO$_2$/TiO$_2$/RuO$_2$ (001) AFMTJs (Fig. 6b) and Fe$_4$GeTe$_2$/vacuum/Fe$_4$GeTe$_2$ lateral AFMTJs [11]. These calculations show that the total field-like torque in the free layer is large, while the total damping-like torque is small, consistent with the expectation of $\tau_A^{FL} = \tau_B^{FL}$ and $\tau_A^{DL} = -\tau_B^{DL}$. The total field-like torque induced by the Néel spin currents is comparable to that in a Fe/MgO/Fe MTJ with similar barrier thickness [104], robust to the interface structure, and thus can be used to generate the ultrafast deterministic switching of the Néel vector in AFMTJs [11].

**Spin torques in noncollinear AFM electrodes**. In a noncollinear antiferromagnet with three magnetic sublattices ($m_A$, $m_B$, and $m_C$) aligned within the Kagome plane, the spin-torque dynamics exhibits rich behavior. For example, when an external spin current is injected into a noncollinear antiferromagnet and its spin polarization $p$ is perpendicular to the Kagome plane, i.e. $p \perp m_\alpha$, the damping-like torque $\tau_\alpha^{DL}$ is uniform. This torque tilts $m_\alpha$ along the out-of-plane direction, causing the finite $\tau_{E,\alpha}^F$ to drive an ultrafast oscillation of $m_\alpha$ within the Kagome plane (like in the case of Fig. 6a, left) [105,106]. In the presence of multi-domain states, such $p$ may drive fast motion of domain walls [107].

If $p$ is in the plane, the resulting torques on the sublattices are different, due to different directions of $m_\alpha$ relative to $p$. For example, in a noncollinear AFMTJ based on a Mn$_3$Pt-type electrode (Fig. 6c) [96], the magnetic group symmetry allows an $x$-directional longitudinal spin current carrying spin polarization along the $y$ direction, i.e. $p \parallel m_C$ [70]. Theoretical modeling shows that this spin current exerts damping-like self-torques on $m_A$ and $m_B$ (Fig. 6c, bottom) [96]. Such self-torques, though interesting, are not able to switch a noncollinear antiferromagnet, because they are internally generated within the antiferromagnet and rotated together with its magnetic order. In contrast, STTs generated by a tunneling spin current injected from another noncollinear AFM electrode can perform the deterministic switching (Fig. 6c, bottom). The self-torques in this case can be useful to reduce the switching current density [96].

Besides the global spin currents, noncollinear local spin currents due to nonrelativistic [96] and relativistic [108] origins can also emerge in noncollinear antiferromagnets. In addition, noncollinear antiferromagnets support the nonrelativistic Rashba-Edelstein effect [109]. These factors can also contribute to spin torques. Furthermore, due to the existence of the nonrelativistic net magnetization [66], many noncollinear antiferromagnets can be considered as weak ferromagnets, and their spin-torque dynamics can be understood as the interplay of weak magnetization and current-induced spin polarization. For example, in an epitaxial noncollinear AFM Mn$_3$Sn with perpendicular net magnetization, SOT switching by a spin Hall current generated from an adjacent heavy metal layer requires an assisting in-plane magnetic field (Fig. 6d) [76-78]. This is like the conventional SOT switching of a perpendicular ferromagnet. Remarkably, a field-free switching of a polycrystalline Mn$_3$Sn film (Fig. 6e) has been reported recently [79], which may combine the different types of spin torques mentioned above.

Finally, we note that STTs in noncollinear AFMTJs are different from those in collinear MTJs [96,110]. The latter cannot occur for the perfect P or AP states, and hence thermal activation is required to induce magnetic fluctuations and activate switching. In a noncollinear AFMTJ, however, STTs can occur in any configuration due to noncollinear sublattice moments (Fig. 6c, bottom) [96]. This eliminates the requirement of thermal activation that suffers from energy dissipation.

## 6. Summary and Outlook

As is evident from this brief review, AFMTJs exhibit interesting functional properties useful for applications. The most notable among them is the giant TMR effect. In conjunction with the possibility of the AFM Néel vector switching by spin torques, it provides potential for novel and more advanced AFM-RAMs. The underlying physics of these properties is determined by the strong exchange interactions in AFM metals, crystal symmetry of the antiferromagnets, and the associated nonrelativistic momentum- and sublattice-dependent spin polarizations. Due to the strong magnetoresistive responses, AFMTJs are expected to be superior to those AFM spintronic devices that are controlled by the relativistically-induced spin-dependent properties associated with a weak SOC [94,95,97,111-113]. However, while AFMTJs have promising perspectives, their investigations are still in a rudimentary stage of development, and thus substantial efforts are required to further elucidate their basic properties and evaluate their value for spintronics.

Although the mechanism of TMR in collinear AFMTJs has been understood, experimental demonstrations are still lacking. This is largely due to difficulties of controlling the Néel vector in collinear antiferromagnets. In this regard, an MTJ with a single FM electrode and an AFM counter electrode can be employed as a preliminary test for using collinear antiferromagnets in AFMTJs [54,55]. In addition, as-grown AFM films usually host a complex domain structure with oppositely aligned AFM domains. Using such films as a reference layer in an AFMTJ would obviously diminish TMR. This problem could be addressed by depositing an AFM film on a FM layer, so that the AFM domains are aligned and switched by an exchange bias [114,115]. Once the domains in the AFM reference layer are well aligned, the Néel vector of the free AFM layer could be deterministically switched by spin torques induced by the Néel spin currents [11].

Ultimately, it is desirable to use electric means to align the collinear AFM domains rather than a magnetic field-controlled exchange bias. Although a spin current with a uniform spin polarization cannot deterministically switch the Néel vector in collinear antiferromagnets on its own, this may be possible with assistance of other factors, such as an external magnetic field, a Dzyaloshinskii-Moriya interaction [116,117], or interfacial uncompensated magnetization. These possibilities are worth further investigations. In addition, in an X-type antiferromagnet, aligning AFM domains is possible by passing an external spin current that exerts a spin torque on a single magnetic sublattice [118]. A similar approach is feasible for AFMTJs with an engineered structure, such as $RuO_2/TiO_2/[TiO_2/CrO_2]_n/CrO_2$ (001) [54].

Spin-torque switching of noncollinear AFM metals have been demonstrated in several experiments [76-79], but requires further investigations. There are various mechanisms to induce spin polarization globally and/or locally in noncollinear antiferromagnets, and various factors affecting its magnitude and direction. Due to relatively low symmetry of noncollinear antiferromagnets, the spin-polarization magnitude is expected to strongly depend on the electric field direction causing dissimilar spin-torque dynamics of noncollinear antiferromagnets with different crystallographic orientations. Net magnetization due to relativistic spin canting [65,66], piezomagnetism [80], and breaking periodicity at the interface [81] may also influence the spin-torque dynamics.

We expect that spin-torque switching of antiferromagnets could be very energy-efficient. To switch the Néel vector, the applied spin torque does not need to directly compete with the strong exchange field, but rather with much smaller magnetic anisotropy and intrinsic damping. Only a slight tilting of the magnetic moments by the spin torque is required to initiate the exchange-driven spin dynamics. Therefore, the associated critical current is expected to be comparable or smaller than that needed for the spin-torque switching of ferromagnets. In addition, the spin-torque switching of antiferromagnets is expected to be much faster than that of ferromagnets. As a result, the comparable or smaller switching current and the much shorter switching time are expected to consume much less energy.

In addition to the spin torque, there are other means to switch the Neel vector in antiferromagnets. For example, it is possible to control the Néel vector by piezoelectric strain induced by electric field [119, 120]. In magnetoelectric antiferromagnets, such $BiFeO_3$ [121,122,123] and $Cr_2O_3$ [124], may be potentially employed as an exchange-coupled under-(over-)layer in AFMTJs to control the Néel vector of the AFM electrode by an electric field applied to the magnetoelectric. The Néel vector may also be switched by voltage controlled magnetic anisotropy (VCMA) [125].

Another issue that needs to be addressed is the material choice for AFMTJs. Among the relatively small number of known altermagnets, only three of them [52,126,127], to the best of our knowledge, are metallic and antiferromagnetic at room temperature. This limits the choice of altermagnetic electrodes for realistic applications. Further material search and design are required [128,129]. In this regard, noncollinear antiferromagnets may be a better choice, because many of them have the Néel temperature above room temperature and two of them, namely $Mn_3Pt$ and $Mn_3Sn$, have already demonstrated their functionality in AFMTJs [9,10]. An important issue which needs to be addressed is magnetic anisotropy which is typically much weaker in noncollinear antiferromagnets than in their collinear counterparts. To realize robust nonvolatile states in noncollinear AFMTJs, optimizing their magnetic anisotropy is necessary. Yet, weak magnetic anisotropy makes noncollinear AFMTJs suitable for magnetic sensor applications [130].

AFMTJs with AFM electrodes that have multiple anisotropy axes may be useful to realize spintronic devices with multiple non-volatile resistance states. For example, in the case on $Mn_3Sn$, different non-volatile resistance states can be obtained associated with the ground-state Néel vector configurations of the AFM electrodes (Fig. 5f). As we have discussed in Sec. 4, switching between these states can be accomplished by a damping-like spin torque.

Tunneling barrier also plays a very important role in the performance of AFMTJs. The widely used MgO in conventional MTJs [22,23] may be not an optimal choice for AFMTJs [9,10]. This is due to the evanescent states in MgO mostly supporting transmission of electrons with the transverse wave vectors $k_∥$ around the center of the 2D Brillouin zone [20,21], where the spin polarization $p_∥$ of the AFM electrodes may be relatively small [6,74]. It would be desirable to search for insulating materials with low decay rates at $k_∥$ away from the zone center to match the spin-polarized conduction channels of AFM electrodes when designing AFMTJs [74]. In addition, if the barrier exhibits a sizable SOC, interesting transport phenomena may occur in AFMTJs due to the interplay between the non-relativistic and relativistic effects, such as unconventional Hall effects [131-133], tunneling anisotropic magnetoresistance [134], and non-reciprocal transport [135].

Finally, probably the most critical requirement for observing the predicted giant TMR effects in AFMTJs is good crystallinity of AFMTJs. Conservation of the transverse momentum $k_∥$ in the process of tunneling imposes stringent conditions on the quality of thin films and heterostructures comprising AFMTJs. Since the momentum- and sublattice-dependent spin polarization in antiferromagnets is highly anisotropic, capabilities for epitaxial growth of AFMTJs along selected directions are required. These challenges need to be addressed by qualified material scientists.

Overall, there are clear indications that AFMTJs can overperform conventional MTJs in terms of the TMR magnitude, switching speed, and packing density. The predicted TMR values are gigantic, and the first experimental data provides a lot of optimism. The switching speed of AFMTJs is expected to be a few orders in magnitude faster. The high packing density is guaranteed by the absence of stray magnetic fields. Thus, AFMTJs have potential to become a new standard for spintronic devices, making this research field rich with opportunities for innovation and new developments.

## Acknowledgments

The authors thank Shui-Sen Zhang and Yuan-Yuan Jiang for their help in preparing figures. This work was supported by the National Science Foundation through DMR (NSF grant No. DMR-2316665) and EPSCoR RII Track-1 (NSF grant OIA-2044049) programs. D.F.S. acknowledges support from the National Key R&D Program of China (Grant No. 2021YFA1600200), the National Natural Science Foundation of China (Grants Nos. 12274411, 12241405, and 52250418), the Basic Research Program of the Chinese Academy of Sciences Based on Major Scientific Infrastructures (Grant No. JZHKYPT-2021-08), and the CAS Project for Young Scientists in Basic Research (Grant No. YSBR-084).